# Reevaluating the Taylor Rule with Machine Learning

## Alper D. Karakas

April 2021

## Abstract

This paper aims to reevaluate the Taylor Rule, through a linear and a nonlinear method, such that its estimated federal funds rates match those actually previously implemented by the Federal Reserve Bank. In the linear method, this paper uses an OLS regression model to find more accurate coefficients within the same Taylor Rule equation in which the dependent variable is the federal funds rate, and the independent variables are the inflation rate, the inflation gap, and the output gap. The intercept in the OLS regression model would capture the constant equilibrium target real interest rate set at 2. The linear OLS method suggests that the Taylor Rule overestimates the output gap and standalone inflation rate's coefficients for the Taylor Rule. The coefficients this paper suggests are shown in equation (2). In the nonlinear method, this paper uses a machine learning system in which the two inputs are the inflation rate and the output gap and the output is the federal funds rate. This system utilizes gradient descent error minimization to create a model that minimizes the error between the estimated federal funds rate and the actual previously implemented federal funds rate. Since the machine learning system allows the model to capture the more realistic nonlinear relationship between the variables, it significantly increases the estimation accuracy as a result. The actual and estimated federal funds rates are almost identical besides three recessions caused by bubble bursts, which the paper addresses in the concluding remarks. Overall, the first method provides theoretical insight while the second suggests a model with improved applicability.

# I. Introduction

The Taylor Rule is a monetary policy tool to assess the target interest rate for a central bank by using monetary policy indicators. These are the real interest rate, the inflation rate gap, and output growth gap. The equation is shown below in equation (1).

(1) $$i_t = \pi_t + r^* + \beta_\pi (\pi_t - \pi_t^*) + \beta_y (y_t - y_t^*)$$

According to John Taylor's papers in 1993 and then in 1999, the target federal funds rate is estimated by setting the $\beta$ coefficients to *0.5*. In 1999, he published a paper that states the coefficient in front of the output gap is merely greater than or equal to zero. However, as shown in *Figure A*, it is clear that the implemented federal funds rate has rarely been even approximately the same as the one estimated by the Taylor Rule — an issue noted in Woodford (2001). There is, then, a question as to whether the federal funds rates the Taylor Rule suggests can be improved. Federal funds rates are a tool to mitigating the financial markets, combating economic growth that is too slow or too fast. Thus, this paper explores if those coefficients should be adjusted in the Taylor Rule equation to better match what has really been implemented by the Federal Reserve Bank (FRB) to maintain macroeconomic equilibrium in the United States.

This paper interprets these coefficients as insight into which aspect of the economy has a greater influence on picking the target interest rate. In the 1993 model both coefficients are *0.5*, so this implies that the inflation gap and the output gap are equally important parameters taken into consideration for picking the target rate. However, with the knowledge that the target rate actually set by the FRB is not identical to that as predicted by the Taylor Rule, exploring the possibility of different coefficients, and thus decision influence, seems necessary. With an understanding of new coefficients, insight, based on the inflation gap and output gap, into where the FRB has allocated focus for setting rates since the 1960s is furthered. Additionally, the Taylor Rule assumes the coefficient in front of the inflation rate term is *1*. However, the inflation rate is still a variable, and therefore its coefficient for targeting a rate is still up for question.

Thus, this paper aims to reevaluate the Taylor Rule, through a linear and a nonlinear method, such that its estimated federal funds rates match those actually previously implemented by the Federal Reserve Bank. In the linear method, this paper uses an OLS regression model to find more accurate coefficients within the same Taylor Rule equation in which the dependent variable is the federal funds rate, and the independent variables are the inflation rate, the inflation gap, and the output gap. The intercept in the OLS regression model would capture the constant equilibrium target real interest rate set at *2*. The linear OLS method suggests that the Taylor Rule overestimates the output gap and standalone inflation rate's coefficients for the Taylor Rule. The coefficients this paper suggests are shown in equation (2). In the nonlinear method, this paper uses a machine learning system in which the two inputs are the inflation rate and the output gap and the output is the federal funds rate. This system utilizes gradient descent error minimization to create a model that minimizes the error between the estimated federal funds rate and the actual previously implemented federal funds rate. Since the machine learning system allows the model to capture the more realistic nonlinear relationship between the variables, it significantly increases the estimation accuracy as a result. The actual and estimated federal funds rates are almost identical besides three recessions caused by bubble bursts, which the paper addresses in the concluding remarks. Overall, the first method provides theoretical insight while the second suggests a model with improved applicability.

(2)
$$i_t = r^* + 0.705\,(\pi_t) + 0.525\,(\pi_t - \pi^*) + 0.13\,(y_t - y_t^*)$$

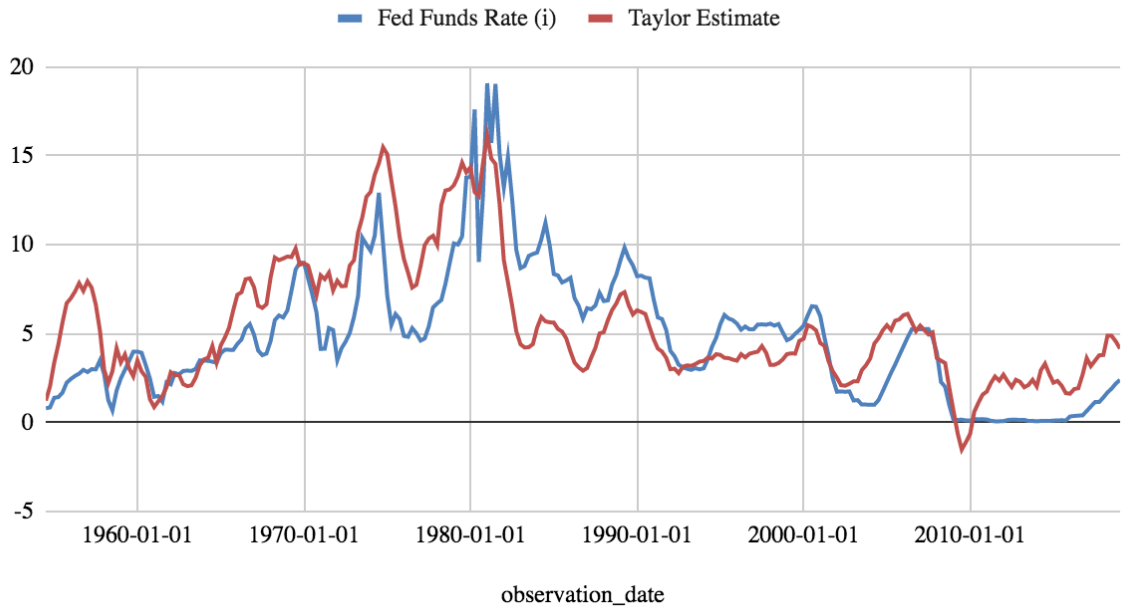

***Figure A***

The rest of this paper is organized as follows. The next section expands the origins of the data. Section III provides background on the Taylor Rule's origins and related literature. Section IV covers the methodologies used to reevaluate the Taylor Rule and summarizes the results. Section V concludes and ponders future research.

## II. Data

As described in the introduction section, the variables needed for the study are the federal funds rate, the inflation rate, the inflation gap, and the output gap. Recall from equation (1), the inflation gap is the difference between the inflation rate at time $t$ ($\pi_t$) and the target inflation rate ($\pi_t^*$) which is set at *2%*. Also, the output gap is the difference between real GDP at time $t$ ($y_t$) and the potential real GDP at time $t$ ($y_t^*$). The last variable necessary is the federal funds rate ($i_t$) actually implemented in year $t$. The equilibrium real interest rate ($r^*$) is set at *2%* like the target inflation rate. All these values — $\pi_t$, $y_t$, $y_t^*$, $i_t$ — come from the FRED St. Louis database which is listed in the References.

## III. Background

In Taylor (1999), it is stated that one monetary policy rule is better than another rule if it results in better economic performance according to some criterion. However, this paper argues that one rule is better than another also if it is more historically realistic because then it can more accurately predict federal funds rate decisions by the Federal Reserve System. By virtue, a better rule uses the same inputs (inflation and output gap) and produces more accurate outputs (federal funds rate). A rule that produces more accurate outputs will more accurately predict future outputs, and therefore, it will guide the direction of the federal funds rate more optimally based on past decisions of the FRB. Essentially, the criterion that makes the rule in this paper better is proximity to past decisions of the FRB. In fact, Taylor (1999) also mentions that a historical approach for monetary policy evaluation is a necessary complement to the model-based approach. A model for setting future federal funds rates is trained by historical data on previous decisions by the FRB. That is exactly what this paper does.

Additionally, in reevaluating the Taylor Rule through a linear and a nonlinear method to estimate a more accurate federal funds rate, this paper does not augment the model with additional inputted variables and parameters which is seen in the STR and STAR model from Castro (2010), Martin and Milas (2004), and Petersen (2007). Through using the same components and assumptions as the simple Taylor Rule, the ML nonlinear model in this paper shows improvement to federal funds rate estimation abilities. The ML model's nonlinearity makes room for any asymmetric preferences and responses in monetary policy.

Furthermore, in Taylor (1993), it is stated that a quarterly time period is too short to average out blips in price level fluctuations, but it is also too long to hold the federal funds rate fixed. Clearly, the frequency of recording interest rate fluctuations impact the analysis. In this paper, however, the federal funds rate change is recorded every quarter since this is how the data is given. Secondly, in Taylor (1993), it is mentioned that of course there are other metrics that observe price level movements, but to stay in line with the inputs of the Taylor Rule, this paper uses the recorded inflation rate as the only metric.

Lastly, like the Taylor Rule, this paper keeps the target inflation rate and the real interest rates fixed at *2%*. Although it is questionable that these values are consistently optimal at *2%,* as the US economic environment evolves over time via natural rates of unemployment, long-run economic growth, and long-run price level growth may evolve with it, since this paper is analyzing the Taylor Rule, the target inflation rate and the real interest rate will also be fixed at *2%*. Furthermore, this paper also assigns a coefficient in front of the solo-standing inflation rate component of the Taylor Rule essentially because it is, too, a random variable whose variation can affect the variation, and hence the result, of the federal funds rate. In fact, the Taylor Rule itself simply uses a coefficient of *1* for the solo-standing inflation rate component.

## IV. Methods & Results

### *IV.I OLS Method*

In the linear method, this paper uses an OLS regression model to find more accurate coefficients within the same Taylor Rule equation. This entails that with the data described previously, the Taylor Rule's equation becomes a model which is empirically tested. This model uses each parameter as a variable. Each variable is a vector of different values given from different points in time ($t$), and, for simplicity, they all are in units of percents rather than the raw decimal form. Therefore, the model used for the OLS method is simply equation (3).

(3) $$i_t = r^* + \beta_1\,(\pi_t) + \beta_\pi\,(\pi_t - \pi^*) + \beta_y\,(y_t - y_t^*)$$

In equation (3), $i_t$ is the federal funds rate — the rate actually implemented, $\pi_t$ is the actual inflation rate for time period $t$, $\pi^*$ is the target inflation rate which is set at *2%*, $y_t$ is the real GDP coined at 2012 dollars for each time period $t$, and $y_t^*$ is the potential real GDP coined at 2012 dollars for each time period $t$. Thus, $(\pi_t - \pi^*)$ is the inflation gap, and $(y_t - y_t^*)$ is the output gap. However, this model is rearranged, shown in (4) to (10), since there is multicollinearity between the inflation gap and inflation as they are not linearly independent.

(4) $$i_t = r^* + \beta_1\,(\pi_t) + \beta_\pi\,(\pi_t) - \beta_\pi(\pi^*) + \beta_y\,(y_t - y_t^*)$$

(5) $$i_t = r^* + (\beta_1 + \beta_\pi)\,\pi_t - \beta_\pi(\pi^*) + \beta_y\,(y_t - y_t^*)$$

Then, in equation (6), $(\beta_1 + \beta_\pi)$ is replaced with $\theta_\pi$.

(6) $$i_t = r^* + \theta_\pi\,(\pi_t) - \beta_\pi(\pi^*) + \beta_y\,(y_t - y_t^*)$$

(7) $$i_t = r^* - \beta_\pi(\pi^*) + \theta_\pi\,(\pi_t) + \beta_y\,(y_t - y_t^*)$$

The value $\beta_\pi$ is still able to found. If we use the given Taylor Rule 1999, where $\beta_\pi$ and $\beta_y$ are both set at *0.5* and $\beta_1$ is set at *1*, then the linear regression where the Taylor predicted interest rate should produce *1* as the constant, $\theta_\pi$ = *1.5*, and $\beta_y$ = *0.5*.

(8) $$\textit{regression intercept } (\alpha) = \ r^* - \beta_\pi(\pi^*) = 2 - 0.5(2) = 1.$$

(9) $$(\beta_1 + \beta_\pi) = \theta_\pi \rightarrow (1 + 0.5) = 1.5.$$

See equations (8) and (9). This is exactly what is produced, as seen in *Table 1* with an $R^2$ value essentially equal to *1*. However, since it is known that $r^*$ and $\pi^*$ are set at *2%*, then the intercept value, $\alpha$, given from the OLS regression results gives the ability to solve for the value $\beta_\pi$. Then, after finding $\beta_\pi$, it is possible to solve for the value $\beta_1$ since the regression results give a value for $\theta_\pi$. Thus equation (10) displays the OLS linear regression model that this paper runs to explore the possibility of different Taylor Rule coefficients when using the implemented federal funds rate as the dependent variable.

(10) $$i_t = r^* - \beta_\pi(\pi^*) + \theta_\pi(\pi_t) + \beta_y(y_t - y_t^*)$$

*Table 1*

| | Taylor Rule Estimated Model Check (1) | Implemented Federal Funds Rate (2) |
|---|---|---|
| *Intercept* ($\alpha$) $r^* - \beta_\pi(\pi^*)$ | 0.999977*** (0.005569) | 0.94731*** (0.26689) |
| *Inflation* ($\pi_t$) | 1.500064*** (0.001406) | 1.22860*** (0.06737) |
| *Output Gap* ($y_t - y_t^*$) | 0.499989*** (0.001344) | 0.13034* (0.06441) |
| *Obs.* | 259 | 259 |
| $R^2$ | 0.9999 | 0.5672 |

With $r^*$ and $\pi^*$ set at *2*, solving for $\beta_\pi$ and $\beta_1$ goes as follows:

(11) $r^* - \beta_\pi(\pi^*) = 2 - \beta_\pi(2) = 0.95 \rightarrow 2 - 0.95 = \beta_\pi(2) \rightarrow 1.05 = \beta_\pi(2) \rightarrow \beta_\pi = 0.525$

(12) $(\beta_1 + \beta_\pi) = \theta_\pi \rightarrow (\beta_1 + 0.525) = 1.23 \rightarrow \beta_1 = 0.705$

Lastly, it is notable that the output gap coefficient is *0.13* rather than *0.5*. Therefore, now that the values for $\beta_1, \beta_\pi,$ and $\beta_y$ are found, the Taylor Rule equation that is more accurately reflective of how the FRB has been behaving since the 1960s is shown in equation (13).

(13) $i_t = r^* + 0.705(\pi_t) + 0.525(\pi_t - \pi^*) + 0.13(y_t - y_t^*)$

These new coefficients suggest, one, the current inflation rate of time $t$ does not have a one-to-one effect on the federal funds rate, rather it is slightly less intense at *0.705*. Two, the inflation gap coefficient set at *0.5* by the 1993 Taylor Rule is quite accurately reflective of the behavior of the FRB since the 1960s found by OLS estimation as it is not substantially different from the *0.525*. Three, the output gap coefficient set at *0.5* by the 1993 Taylor Rule is not the most accurate reflection of the FRB's behavior because this paper finds that the influence the output gap has on the implemented federal funds rate is just *0.13*. This value is substantially different from *0.5*. Furthermore, when comparing the influence the inflation and output gap seem to have on the implemented federal funds rate, the difference between *0.525* and *0.13* implies that an inflation rate gap has a higher weight, or a higher influence, on the federal funds rate that the FRB implements. This implication from the reevaluated coefficients is in line with the relative influences found in Castro (2010), Martin and Milas (2004), and Petersen (2007). In the 1993 Taylor Rule equation, these weights are set as the same, meaning that both gap parameters have equal influence on the value the rate is set at; however, based on the actual rates set, this paper suggests that this is not true. Additionally, the intercept and the inflation gap variable coefficient are both significant at the *0.1%* level, whereas the coefficient for the output gap is significant only at the *5%* level. Although the output gap coefficient is still significant, the disparity in the level of significance implies even further that the inflation gap has more influence than the output gap.

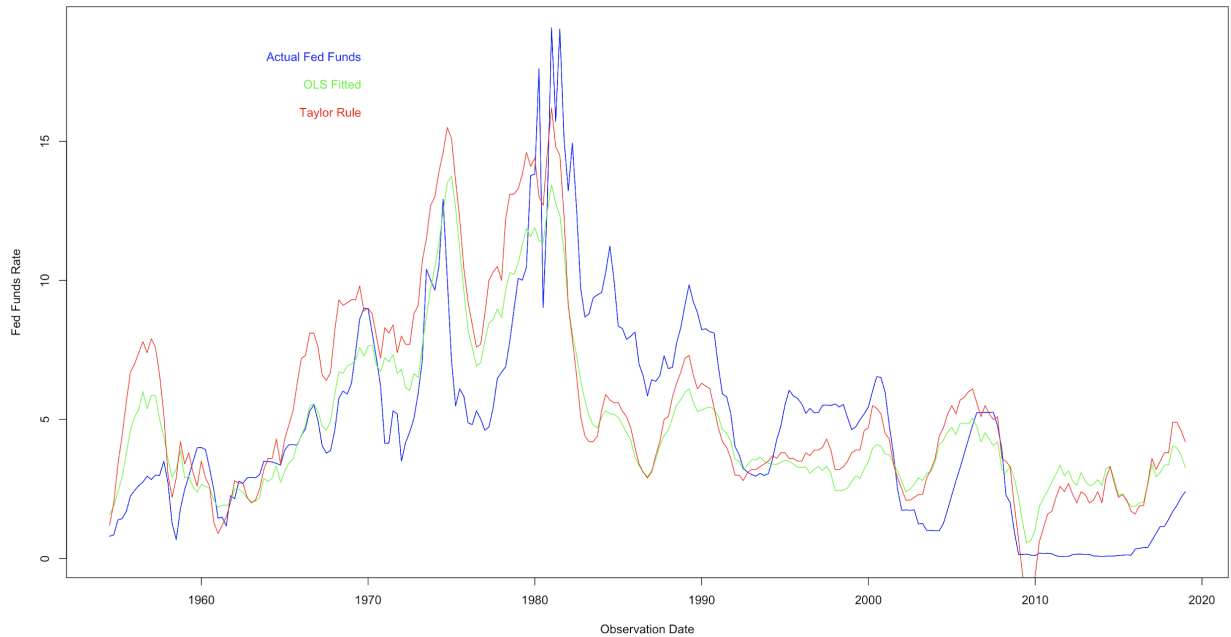

***Figure B***

    *Figure B* shows the estimation strength of the Taylor Rule, like in *Figure A*, as well as

that found through an OLS fit. Again, it is observable that although the Taylor Rule's estimate

does a good job at suggesting the direction of the federal funds rate, there are noticeable gaps in

the actual values of the rate. The OLS estimation reevaluates the Taylor Rule's coefficients

attempting to mitigate this issue. In doing so, the OLS fit also accurately suggests the direction of

the federal funds rate based on inflation and output gap, and it outputs a rate value more

accurately than the Taylor Rule. The sum of the residuals for the OLS fit is *2.81 • e$^{-8}$* while that

of the Taylor Rule is *-169.49*. However, the absolute value sum of the residuals for the OLS fit is

*510.54* while that of the Taylor Rule is *571.25*. The rule with redefined coefficients does produce rate values closer to those that were actually implemented, but the OLS fit's estimates are not significantly more accurate than the Taylor Rule's. In order to pursue the line-fitting goal of diminishing the gap between an estimated value and the actual federal funds rate, this paper uses a machine learning method.

<u>*IV.II Machine Learning Method*</u>

At the heart of the nonlinear, machine learning (ML) method, the essential solution it offers is modeling a system that occurs in reality. For the case of this paper, the actual system the ML-based approach seeks to model is the one that connects the inflation rate and the output gap to the federal funds rate, like the Taylor Rule. The assumption is that the system that connects real life inputs and outputs is a nonlinear function — a relationship that is much more complex than the Taylor Rule equation suggests. In fact, Petersen (2007) finds that the FRB follows a nonlinear Taylor Rule, not a linear one. The goal is to "train" a neural network system that behaves like the one that plays out in reality. *Figure C* delineates this goal: the estimated federal funds rate at time *t* should equal the actual federal funds rate at time *t* for the full vector of all *t* intervals. The ML method allows a neural network to automatically model the relationship between the two inputs and the output.

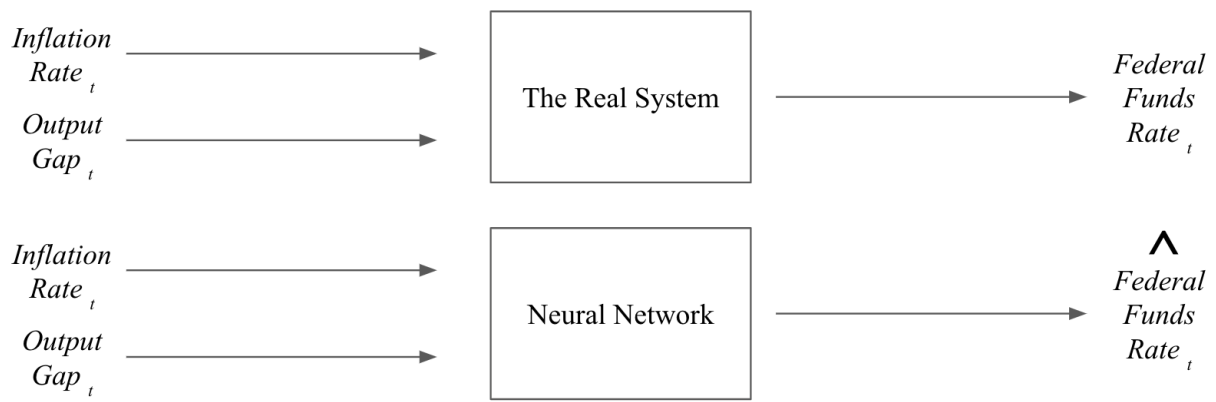

**Figure C**

Thus, the result is a function, $\boldsymbol{f}$, of the inflation rate and the output gap which can be summarized as $\hat{i}_t = \boldsymbol{f}(\pi_t, (y_t - y_t^*))$ where, at time interval $t$, $\hat{i}_t$ is the estimated federal funds rate, $\pi_t$ is the inflation rate, and $(y_t - y_t^*)$ is the output gap. The "training" process ML undergoes happens by continuously comparing the estimated federal funds rate to the actual one for each time interval and adjusting the weights in the function until the estimated value comes extremely close to the actual one. Mathematically, this proximity between the two values can be written as, $\hat{i}_t - i_t = e$ where $e < (\varepsilon = 10^{-4})$. This training process is elaborated further in this section.

First, it is important to discuss the calculation process from the inputs to the output, which *Figure D* and *Figure E* outline. Beginning with the inputs, there are three layers that are marked by nodes in this example with one hidden layer (Rabiner 1989). The hidden layer is made up of six nodes ($N = 6$) in this example. At each node ($n = 1, 2, \ldots 6$), the input values, $j = \pi, (y - y^*)$, are received through a function that multiplies it with a weight, $w_{jn}$, which then becomes a new value, $u_n$, and can be understood in equation (14). Then the node in the hidden layer converts the received value into a sigmoidal function, $z_n$, and sends a value to the output layer. This sigmoidal function can be understood in equation (15). By using a sigmoid function, not only does the system become nonlinear, but the sigmoid function allows the model to become continuous and bounded between *1* and *0*. A bounded and continuous function is important for the ML system's training process which, again, is elaborated later in this section (Plaunt, et al. 1986). Finally, the node at the output layer receives the $z_n$ value and multiplies it with another weight, $v_n$, which then becomes the estimated federal funds rate $\hat{i}_t$. This result can be understood in equation (16). Thus, at the end of the calculation process, the model produces a $\hat{i}_t$. Now, the training component comes in when comparing the $\hat{i}_t$ value to $i_t$.

(14) $$u_n = (w_{\pi n} \cdot \pi_t) + (w_{(y \text{-} y^*)n} \cdot (y \text{-} y^*)_t)$$

(15) $$z_n(u_n) = \frac{1}{1 + e^{-u}}, \quad \text{where } u = u_n$$

(16) $$\hat{i}_t = \sum_{n=1}^{6} v_n + z_n$$

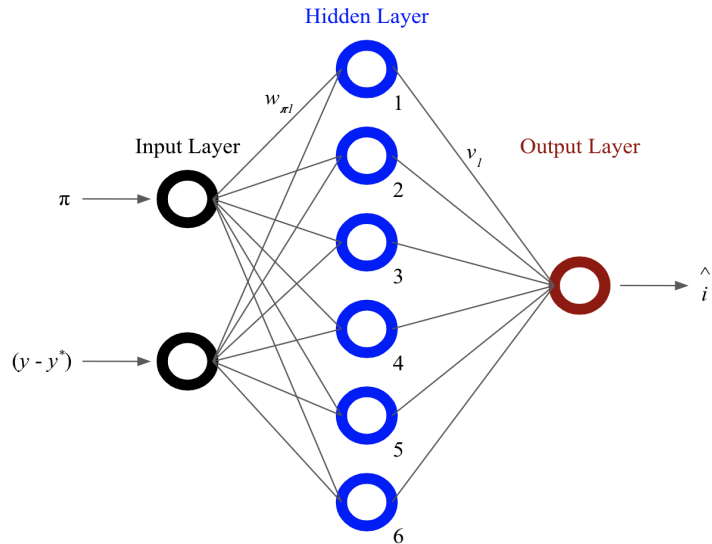

**Figure D**

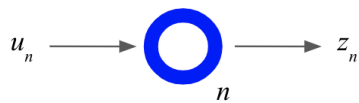

**Figure E**

Essentially there are two passes. In the first pass, which goes forward from input data to output data, the values $u_n$, $z_n$, and ultimately $\hat{i}_t$ are calculated using $w_{jn}$ and $v_{jn}$. The second pass goes backwards; it is known as backpropagation. Given the error $(i_t \text{-} \hat{i}_t = e)$, the weights $w_{jn}$ and $v_{jn}$ are adjusted by using a gradient descent optimization method, since the goal is to get $\hat{i}_t$ as close to $i_i$ as possible. With each weight going through its own updating process, there is an ongoing loop of trial and error to find the perfect weights. Values for weights are attempted at every trial iteration; then, after error is calculated, new values for weights are attempted again with the goal of reducing error.

Therefore error changes as the weights do, which is represented as $\frac{\partial E}{\partial w}$. A fixed step size, $\mu$, is multiplied by $\frac{\partial E}{\partial w}$ to produce the amount that the weights are adjusted. Imagine weights as knobs that are adjusted by increments of $\mu(\frac{\partial E}{\partial w})$ until error hits its minimum. In all, at every increment of $w_{jn}$, a new error value, $e$, is produced, and for every value $e$ can be plotted on the quadratic error function, $E(e)$ (Plaunt, et al. 1986). Thus, as $e$ converges to its minimum, there is a convergence to the vertex of the error function $E(e)$, as depicted in *Figure F*.

Here, it is important to acknowledge the sigmoid function's importance, $z_n$. By using a sigmoid function, the system becomes nonlinear, bound between *1* and *0*, and continuous. The nonlinearity allows for modeling complex relationships. A bounded function in the hidden layer gives the gradient descent process direction towards the minimum of the error function so that the risk of diverging from the minimum error rather than converging is absent. Lastly, its continuity allows the gradient descent process to differentiate the error function ($\frac{\partial E}{\partial w}, \frac{\partial E}{\partial v}$) at any given segment of the error function.

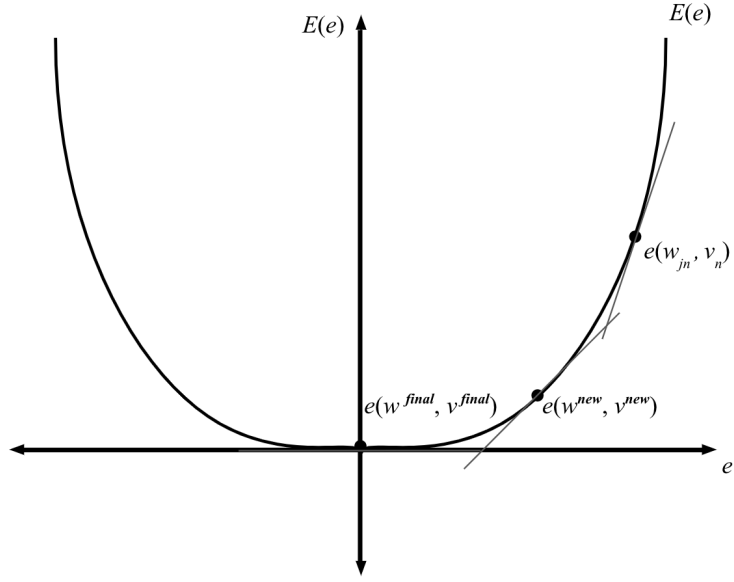

**Figure F**

Note, $E$ is a function of error, $e$, which is defined in equation (17). It is set as a positive parabola which allows it to be differentiable during the weight update process and also allows $e$ to reach a minimum value. To begin with weight $w_{jn}$, equations (18) to (24) outline its update process, where $\mu$ is a fixed step size.

(17) $$E(e) = \frac{1}{2}e^2 = \frac{1}{2}(i - \hat{i})^2$$

(18) $$w^{new} = w^{old} - \mu\left(\frac{\partial E}{\partial w}\right)$$

(19) $$\frac{\partial E}{\partial w} = \frac{\partial E}{\partial \hat{i}} \cdot \frac{\partial \hat{i}}{\partial z} \cdot \frac{\partial z}{\partial u} \cdot \frac{\partial u}{\partial w}$$

(20) $$\frac{\partial E}{\partial \hat{i}} = -e$$

(21) $$\frac{\partial \hat{i}}{\partial z} = v_n$$

(22)
$$\frac{\partial u}{\partial w} = \pi$$

(23)
$$\frac{\partial z}{\partial u} = \partial \frac{1}{1+e^{-u}} \cdot \partial u = \frac{1}{1+e^{-u}} \cdot (1 - \frac{1}{1+e^{-u}}) = z_\pi \cdot (1 - z_\pi)$$

(24)
$$\therefore w^{new} = w^{old} + \mu(ev_n\pi z_\pi \cdot (1 - z_\pi))$$

Equations (25) through (30) explain how the weight after the hidden layer, $v_n$, is updated where $\mu$ is a fixed step size.

(25)
$$E(e) = \frac{1}{2}e^2 = \frac{1}{2}(i - \hat{i})^2$$

(26)
$$v^{new} = v^{old} - \mu(\frac{\partial E}{\partial v})$$

(27)
$$\frac{\partial E}{\partial v} = \frac{\partial E}{\partial \hat{i}} \cdot \frac{\partial \hat{i}}{\partial v}$$

(28)
$$\frac{\partial \hat{i}}{\partial v} = z_\pi$$

(29)
$$\frac{\partial E}{\partial v} 2 \cdot \frac{1}{2}(i - \hat{i}) \cdot \frac{\partial \hat{i}}{\partial v} = -ez_\pi$$

(30)
$$\therefore v^{new} = v^{old} + \mu(ez_\pi)$$

Thus, machine learning through gradient descent is an effective tool to minimize error between a model and the true occurrence. In the case of this paper, the model is a function of the output gap and inflation rate to output a federal funds rate, and the true occurrence is the actual federal funds rate enacted at observation date $t$. The goal, then, is to match the model to the true occurrence with minimal difference. As seen in *Figure G*, by only using the two inputs, the ML GD method produces a model that mirrors the true used federal funds rate observed at date $t$ significantly better than that of the Taylor Rule and the OLS fit. It is important to flag that there are three times where the ML estimated $\hat{i}_t$ is significantly different from $i_t$. These three quarters are during the recessions in 1984, 2001, and 2009, or the Energy Crisis, Dot-Com Bubble, and Housing Bubble, respectively. Specific bubbles have their own idiosyncrasies which makes producing a general model more difficult. However as a future scope, the ML system can be altered by incorporating a "bubble burst" recession dummy as the third input. This input entails that if at time interval $t$, a bubble burst and led to a recession, then the input's value is $1$. If at time interval $t$, there is no such bubble burst — even if the business cycle enters a recessionary phase — then this input's value is $0$. This input is excluded from this paper because its goal is to reevaluate the Taylor Rule with the OLS and ML methods. The Taylor Rule has no such input: its inputs are only inflation and the output gap. The figure still shows the improvement ML provides from the Taylor Rule. With this improved model, forecasts and projects can be estimated more accurately.

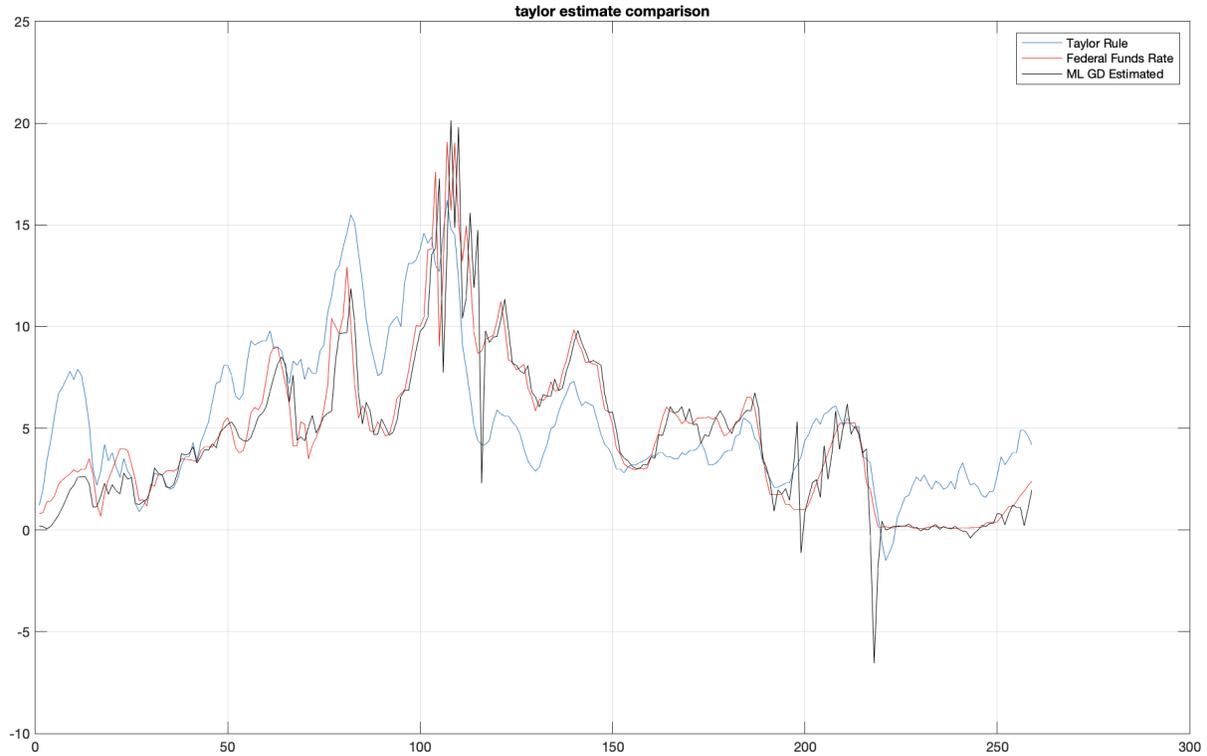

*Figure G*

## V. Concluding Remarks

This paper reevaluates the monetary policy rules defined in Taylor (1993) and Taylor

(1999) in a linear and a nonlinear manner. First, through an OLS method, the value of the

coefficients in the Taylor Rule are questioned. By running an OLS regression on the Taylor Rule

equation and using the actually implemented federal funds rate as the dependent variable, the

Taylor Rule's coefficients are redefined and output federal funds rate estimates that are

noticeably but not significantly more accurate. The OLS fitted estimates decreased the absolute

value sum of residuals from *510.54* to *571.25* and the calculated sum of residuals from *-169.49*

to *2.81 • e^{-8}*. Equations (20) and (21) display how the coefficients changed when running an OLS

regression. These new coefficients suggest that the output gap and the standalone inflation rate's weight on the federal funds rate is overestimated in Taylor (1999).

(31)      *1999 Taylor Rule*: $i_t = \pi_t + r^* + 0.5\,(\pi_t - \pi_t^*) + 0.5\,(y_t - y_t^*)$

(32)      *OLS Fitted Taylor Rule*: $i_t = r^* + 0.705\,(\pi_t) + 0.525\,(\pi_t - \pi^*) + 0.13\,(y_t - y_t^*)$

The nonlinear manner evaluates the Taylor Rule in its accuracy to predict the federal funds rate with inflation and output gap as the only inputs. It is observed that the Taylor Rule and the OLS fitted rule produce estimated rates that are noticeably different from the rate actually implemented because they assume a linear relationship between the variables. Thus, this paper executes a line-fitting exercise to minimize the error between estimation and truth through a gradient descent machine learning method. This nonlinear method estimates a federal funds rate that is significantly more accurate than the Taylor Rule or the OLS regression by only using inflation rate and output gap as inputs. Therefore, being trained with historical data, this GD ML model can predict future federal funds rates significantly more accurately than the Taylor Rule can by using the same two inputs.

It is important to note that the OLS method reevaluates the Taylor Rule with a theoretical lens while the GD ML method does so through an application lens. In the OLS method, this paper uses the same framework as the Taylor Rule to examine a more accurate weight of the independent variables. Rather than seeking the perfect federal funds rate, the OLS method finds the degree to which the output gap and the inflation gap seem to influence the federal funds rate that was implemented. The GD ML method focuses on the ability to predict an accurate rate rather than the relationships between the independent and dependent variables. Therefore, it is more effective in application but less explicitly descriptive for the individual roles each input plays.

Further research is contemplated in two ways. First, the ML system's accuracy increases as it utilizes more inputs. Increasing the amount of inputs would go beyond the Taylor Rule's scope, by definition, but seeing how accurately the system can model set federal funds rates over time with only two inputs, improving the accuracy even more with additional inputs implies this tool's modeling power. An example of utilizing more inputs to increase the model's accuracy is the "Bubble Burst" recession dummy variable. Including this additional input would take into account the FRB's unique behavior when recessions are caused by bursted bubbles such as the dot-com or housing bubbles. Beyond just this input, others can include various exchange rates, change in trade tariffs, or the unemployment rate.

Secondly, the ML system can be revised to obtain forecasting or predictive abilities. Currently, at inputs and output being from time $t$, the system offers the ability to simulate hypothetical events. Rather than the inputs and the output being of the same time interval $t$, the inputs would be taken from time $t-1$ while the output still being from time $t$. This adjustment would not be pragmatic with the OLS method because its model is fixed to the Taylor Rule equation, which assumes the same time interval for inputs and outputs. However, the ML method has the ability to adapt its model to a differing dataset, which underscores ML's modeling power. Therefore, by training the ML model to replicate the nonlinear system between the past inflation rate and output gap and the current federal funds rate, the system is able to take the current inflation rate and output gap (at time $t$) and then predict a future federal funds rate (at time $t+1$). Using the logic from the first idea for further research scopes, an increase of inputs will increase the forecasting accuracy. This paper implies that utilizing a ML method to model real world economic systems and relationships increases the model's accuracy and applicability. In all, this accurate, simple, and efficient method to model systems in the real world holds great usability in economic research.